\documentclass{article}

\usepackage{microtype}
\usepackage{graphicx}
\usepackage{subcaption}
\usepackage{booktabs} 
\usepackage{xurl}

\usepackage{hyperref}

\usepackage[preprint]{icml2026}

\usepackage{amsmath}
\usepackage{amssymb}
\usepackage{mathtools}
\usepackage{amsthm}

\usepackage[capitalize,noabbrev]{cleveref}
\usepackage{array}
\usepackage{booktabs}
\usepackage{graphicx}
\usepackage[table]{xcolor}

\theoremstyle{plain}

\theoremstyle{definition}

\theoremstyle{remark}

\usepackage[textsize=tiny]{todonotes}

\icmltitlerunning{The Closing Window: How Governments Could Lose Their Ability to Restrain Advanced AI}

\begin{document}

\twocolumn[
  \icmltitle{The Closing Window:\\How Governments Could Lose Their Ability to Restrain Advanced AI}

  \icmlsetsymbol{equal}{*}

  \begin{icmlauthorlist}
    \icmlauthor{Peter Barnett}{sch}
  \end{icmlauthorlist}

  \icmlaffiliation{sch}{Machine Intelligence Research Institute, Berkeley, USA}
  \icmlcorrespondingauthor{Peter Barnett}{peter@intelligence.org}

  \icmlkeywords{AI Governance, Machine Learning, ICML}

  \vskip 0.3in
]

\printAffiliationsAndNotice{}

\begin{abstract}
As AI capabilities advance, AI systems will pose greater risks to national security and potentially humanity as a whole. Governments may eventually conclude that these risks warrant restraining AI development. This motivates the question: will governments still be able to restrain AI development in the future, should they want to do so? In this paper we analyze which world events and changes to the state of AI development would make future governance more difficult or even effectively impossible. Our analysis surfaces likely pathways that would lead to these difficulties, including hardware proliferation, continued algorithmic progress, and the release of catastrophically dangerous AI models. Due to the field’s lack of understanding of AI development, it may be difficult or impossible to know when we will hit a “point of no return”, and we therefore recommend a conservative approach. The window may be closing, but governments currently have an opportunity to preserve their optionality if they act soon. Our policy recommendations would enable governments to restrain AI development in the future, while imposing relatively small costs today. 
\end{abstract}

\section{Introduction}

AI capabilities have increased rapidly in recent years, with no clear sign of slowing~\citep{ho2025rosetta,kwa2025measuring}. As AI systems grow more powerful, they may pose a host of risks to society~\citep{bengio2024managing,zelikow2024defense, PE-A3691-4}, including the risk of human extinction~\citep{cais_statement_2023}. Governments may eventually conclude that such risks warrant major restrictions on further AI development and deployment~\citep{barnett2025technical}. Even if they do not think these restrictions are justified today, it is valuable to preserve this option in case they are desired in the future. In this paper, we ask: will it still be feasible to apply these restraints?

It is not only AI systems directly capable of catastrophic or extinction-level harm that merit restraint. Weaker systems, such as those that can accelerate AI development, may develop catastrophically dangerous successors, and so also merit restraint~\citep{chan2026measuring, davidson2025threetypesofinte, field2026ai}. This also includes systems that, while not themselves immediately dangerous, could be modified either to help produce catastrophically harmful AI systems, or to become catastrophically harmful themselves. We will refer to systems that meet any of these criteria (can cause direct catastrophic harm; can substantially aid in the development of directly harmful systems; can be modified to cause direct harm; can be modified to aid in the development of directly harmful systems) as “catastrophically dangerous”. A regulatory regime that permits the development or release of any catastrophically dangerous AI system may fail to preserve the option of future restraint.

A successful governance regime most likely prevents the \textit{training} of such systems. It is prohibitively difficult to verify the comprehensive deletion of a catastrophically dangerous model once it is trained, as we discuss in \cref{sec:dangerousmodel_dev}. Containing a catastrophically dangerous model would require extending restrictions to inference hardware that could run the model, a substantially harder task than restricting training. A frontier model may require 10,000 or 100,000 GPUs to train, but that same model can then be run on a single cluster of only a few GPUs~\citep{epoch2023tradingoffcomputeintrainingandinference, epoch2025trainoncedeploymanyaiandincreasingreturns}. This is especially important for models capable of automating AI research, since such models could drive rapid algorithmic progress through inference alone. This sharp distinction between training and inference reflects the current AI paradigm, and future developments might blur the line. 

\begin{figure*}[t!]
  \centering
  \includegraphics[width=\textwidth]{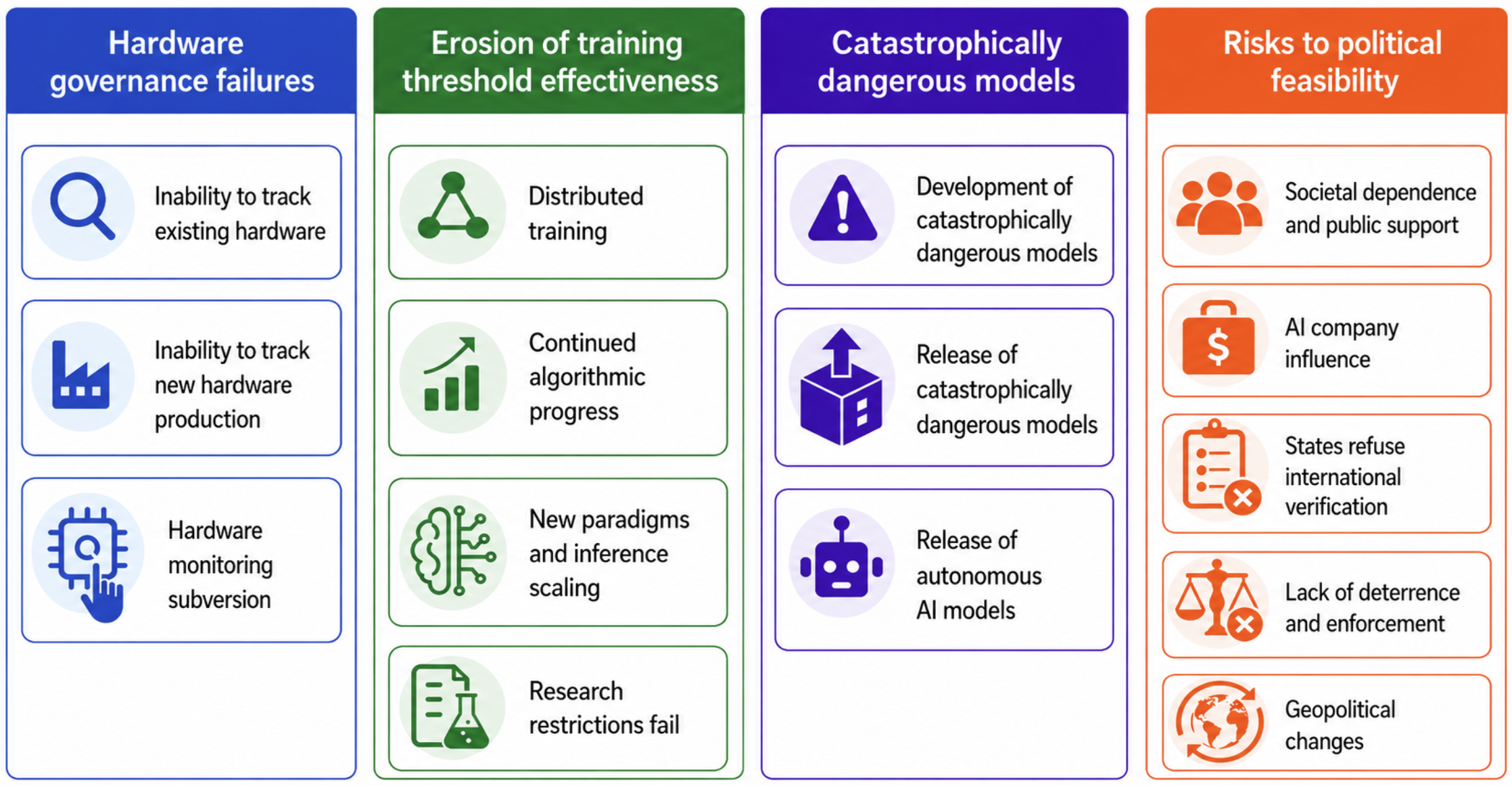}
  \caption{Summary of the pathways to losing the option to restrain AI
development}
  \label{fig:summary}
\end{figure*}

In this paper, we describe a plan for restraining AI development and, based on this, we lay out pathways via which governments would lose the option to restrain AI development, as summarized in Figure~\ref{fig:summary}. We end with a discussion of policies that could help governments preserve this ability. 
\vspace{-0.5cm}
\section{A plan for restraining AI development}

This paper takes preserving the option to restrain AI development to mean the following: \textit{If major governments came to believe that further frontier AI development posed unacceptable risks, they would retain the capacity to impose sufficient restrictions.} 

We consider a regime built around training compute thresholds, based on the proposed international agreement in~\citet{scher2025international}. In this regime, states prohibit training above a specified training compute threshold, restrict research that would reduce the efficacy of this threshold, and verify compliance of other states. The key elements are:

\begin{itemize}
\item Training runs exceeding a specified compute threshold are prohibited.   
\item To facilitate this prohibition, all clusters of AI chips above a certain size are consolidated into monitored data centers.   
\item Hardware monitoring mechanisms verify that chips are not used for prohibited training, and guard against tampering.   
\item AI chip production is closely tracked to ensure that only monitored data centers receive new chips.   
\item Research that would undermine this regime, such as research increasing general-purpose AI capabilities or advancing distributed training, is restricted.   
\item The above measures are implemented domestically and subject to international verification and monitoring, so that states can be confident that potential rivals are also complying.
\end{itemize}

Under this regime, inference on existing AI systems would be permitted, since these systems are not catastrophically dangerous. If catastrophically dangerous AI systems were created, inference would require monitoring, either by detecting prohibited workloads (such as biological weapons development or AI R\&D), or by verifying that chips run only approved, non-dangerous models~\citep{sharma2025constitutional, karvonen2025difr, rinberg2025verifying}. This would also require consolidating and monitoring even small AI clusters capable of running catastrophically dangerous models. As such, containing dangerous capabilities in such a situation may prove prohibitively difficult; it would be far less costly to avoid training dangerous models at all.

\subsection{Connection with broader governance goals}

Our analysis is relevant beyond the specific scenario of a full halt on AI capabilities advancement. Many AI governance objectives rely on having the ability to monitor AI chip use~\citep{heim_governing_2024, sastry_computing_2024, baker2025verifying}. Restraining AI development can encompass a range of actions, including slowing the pace of development or steering it in particular directions---for example, ensuring that AI systems are trained using certain safety techniques, or that certain categories of research proceed more cautiously. A state pursuing any of these must be able to monitor training within its borders, especially if they require high assurance. The governance infrastructure required to steer or slow AI development is largely the same as the infrastructure required to halt it.

Even for a state confident in its domestic AI governance, international economic and military pressures may create incentives to pursue more advanced AI systems. A state that unilaterally slows its own AI development risks falling behind rivals. Sustained restraint of almost any kind thus implies international verification and mutual assurance.

These measures are less useful for governments willing to accept considerably lower levels of assurance and therefore lower levels of safety. For example, comprehensive monitoring of training runs may not be directly applicable to governance regimes based primarily on legal liability or on reporting of results from AI company safety frameworks.

\section{Pathways to losing the option to restrain AI development}

States might forfeit the option to restrain AI development in several ways. Some ways make restrictions more costly or intrusive; others make them harder to enforce or verify, potentially to the point where such restraint becomes infeasible. We group these pathways into four categories: hardware governance failures, erosion of training threshold effectiveness, catastrophically dangerous model development, and risks to political feasibility. These pathways are not independent; algorithmic progress, for example, can weaken compute thresholds and thereby necessitate monitoring smaller clusters. 

Table~\ref{tab:pathway-dependencies} provides an overview of how the pathways interact with each other; pathways may directly cause others or exacerbate the effects of others. 

\definecolor{catH}{HTML}{1F6FB2}  
\definecolor{catE}{HTML}{2E8B57}  
\definecolor{catD}{HTML}{6A4C93}  
\definecolor{catP}{HTML}{D9822B}  
\definecolor{cD}{HTML}{F0C8BC}
\definecolor{cW}{HTML}{F3DDB0}
\definecolor{cDiag}{HTML}{E6E6E6}
\definecolor{rev1}{HTML}{E2937E}
\definecolor{rev2}{HTML}{F2C088}
\definecolor{rev3}{HTML}{FBEED0}
\definecolor{gridgray}{HTML}{B8B8B8}
\newcolumntype{Y}{>{\centering\arraybackslash}m{1.5em}}
\newcommand{\nbox}[1]{\parbox[c][1.7\baselineskip][c]{3.55cm}{\scriptsize #1}}

\begin{table*}[t]\centering\footnotesize
\setlength{\tabcolsep}{2pt}\setlength{\arrayrulewidth}{0.8pt}
\begin{tabular}{@{}!{\color{black}\vrule}>{\centering\arraybackslash}m{1.5em}!{\color{gridgray}\vrule width 0.8pt}m{3.7cm}!{\color{gridgray}\vrule width 0.8pt}>{\centering\arraybackslash}m{2.05cm}!{\color{black}\vrule width 1pt}Y!{\color{catH}\vrule width 0.8pt}Y!{\color{catH}\vrule width 0.8pt}Y!{\color{catE}\vrule width 0.8pt}Y!{\color{catE}\vrule width 0.8pt}Y!{\color{catE}\vrule width 0.8pt}Y!{\color{catE}\vrule width 0.8pt}Y!{\color{catD}\vrule width 0.8pt}Y!{\color{catD}\vrule width 0.8pt}Y!{\color{catD}\vrule width 0.8pt}Y!{\color{catP}\vrule width 0.8pt}Y!{\color{catP}\vrule width 0.8pt}Y!{\color{catP}\vrule width 0.8pt}Y!{\color{catP}\vrule width 0.8pt}Y!{\color{catP}\vrule width 0.8pt}Y!{\color{black}\vrule width 0.8pt}@{}}
\arrayrulecolor{black}\hline
\rule{0pt}{3\baselineskip} & &  & \multicolumn{3}{c}{\textcolor{catH}{\itshape\shortstack{Hardware\\governance\mbox{}}}} & \multicolumn{4}{c}{\textcolor{catE}{\itshape\shortstack{Threshold\\erosion}}} & \multicolumn{3}{c}{\textcolor{catD}{\itshape\shortstack{Dangerous\\models}}} & \multicolumn{5}{c!{\color{black}\vrule width 0.8pt}}{\textcolor{catP}{\itshape\shortstack{Political\\feasibility}}} \\[2pt]\noalign{\vskip -10pt}
 & \textbf{Pathway} & \textbf{Reversibility} & \textcolor{catH}{H1} & \textcolor{catH}{H2} & \textcolor{catH}{H3} & \textcolor{catE}{E1} & \textcolor{catE}{E2} & \textcolor{catE}{E3} & \textcolor{catE}{E4} & \textcolor{catD}{D1} & \textcolor{catD}{D2} & \textcolor{catD}{D3} & \textcolor{catP}{P1} & \textcolor{catP}{P2} & \textcolor{catP}{P3} & \textcolor{catP}{P4} & \textcolor{catP}{P5} \\
\arrayrulecolor{black}\hline
\textcolor{catH}{H1} & \nbox{Inability to track existing hardware} & \cellcolor{rev2}{\scriptsize Difficult} & \cellcolor{cDiag} &  &  & \cellcolor{cW}\textbf{E} & \cellcolor{cW}\textbf{E} &  & \cellcolor{cW}\textbf{E} & \cellcolor{cD}\textbf{C} &  &  &  &  & \cellcolor{cW}\textbf{E} & \cellcolor{cW}\textbf{E} &  \\
\arrayrulecolor{catH}\hline
\textcolor{catH}{H2} & \nbox{Inability to track new hardware\\production} & \cellcolor{rev2}{\scriptsize Difficult} &  & \cellcolor{cDiag} &  & \cellcolor{cW}\textbf{E} & \cellcolor{cW}\textbf{E} &  & \cellcolor{cW}\textbf{E} & \cellcolor{cD}\textbf{C} &  &  &  &  & \cellcolor{cW}\textbf{E} & \cellcolor{cW}\textbf{E} &  \\
\arrayrulecolor{catH}\hline
\textcolor{catH}{H3} & \nbox{Hardware monitoring subversion} & \cellcolor{rev3}{\scriptsize Reversible} &  &  & \cellcolor{cDiag} & \cellcolor{cW}\textbf{E} &  &  &  & \cellcolor{cD}\textbf{C} &  &  &  &  & \cellcolor{cD}\textbf{C} &  &  \\
\arrayrulecolor{catH}\hline
\textcolor{catE}{E1} & \nbox{Distributed training} & \cellcolor{rev2}{\scriptsize Difficult} &  &  &  & \cellcolor{cDiag} & \cellcolor{cW}\textbf{E} &  &  & \cellcolor{cD}\textbf{C} & \cellcolor{cW}\textbf{E} &  &  &  & \cellcolor{cW}\textbf{E} &  &  \\
\arrayrulecolor{catE}\hline
\textcolor{catE}{E2} & \nbox{Continued algorithmic progress} & \cellcolor{rev1}{\scriptsize Almost Impossible} &  &  &  & \cellcolor{cW}\textbf{E} & \cellcolor{cDiag} &  &  & \cellcolor{cD}\textbf{C} &  &  &  &  &  &  &  \\
\arrayrulecolor{catE}\hline
\textcolor{catE}{E3} & \nbox{New paradigms and inference\\scaling} & \cellcolor{rev1}{\scriptsize Almost Impossible} &  &  &  &  &  & \cellcolor{cDiag} &  & \cellcolor{cD}\textbf{C} &  &  &  &  & \cellcolor{cW}\textbf{E} &  &  \\
\arrayrulecolor{catE}\hline
\textcolor{catE}{E4} & \nbox{Research restrictions fail} & \cellcolor{rev2}{\scriptsize Difficult} &  &  &  & \cellcolor{cD}\textbf{C} & \cellcolor{cD}\textbf{C} & \cellcolor{cD}\textbf{C} & \cellcolor{cDiag} &  &  &  &  &  &  &  &  \\
\arrayrulecolor{catE}\hline
\textcolor{catD}{D1} & \nbox{Development of catastrophically\\dangerous models} & \cellcolor{rev1}{\scriptsize Almost Impossible} &  &  &  &  &  &  & \cellcolor{cW}\textbf{E} & \cellcolor{cDiag} & \cellcolor{cD}\textbf{C} & \cellcolor{cD}\textbf{C} &  &  &  &  &  \\
\arrayrulecolor{catD}\hline
\textcolor{catD}{D2} & \nbox{Release of catastrophically\\dangerous models} & \cellcolor{rev1}{\scriptsize Almost Impossible} &  &  &  &  &  &  & \cellcolor{cW}\textbf{E} &  & \cellcolor{cDiag} &  &  &  &  &  &  \\
\arrayrulecolor{catD}\hline
\textcolor{catD}{D3} & \nbox{Release of autonomous AI models} & \cellcolor{rev1}{\scriptsize Almost Impossible} &  &  &  &  &  &  &  &  &  & \cellcolor{cDiag} &  &  &  &  &  \\
\arrayrulecolor{catD}\hline
\textcolor{catP}{P1} & \nbox{Societal dependence and public\\support} & \cellcolor{rev2}{\scriptsize Difficult} &  &  &  &  &  &  &  & \cellcolor{cW}\textbf{E} & \cellcolor{cW}\textbf{E} &  & \cellcolor{cDiag} & \cellcolor{cW}\textbf{E} &  &  &  \\
\arrayrulecolor{catP}\hline
\textcolor{catP}{P2} & \nbox{AI company influence} & \cellcolor{rev3}{\scriptsize Reversible} &  &  &  &  &  &  &  &  &  &  & \cellcolor{cW}\textbf{E} & \cellcolor{cDiag} & \cellcolor{cW}\textbf{E} &  &  \\
\arrayrulecolor{catP}\hline
\textcolor{catP}{P3} & \nbox{States refuse international\\verification} & \cellcolor{rev3}{\scriptsize Reversible} & \cellcolor{cW}\textbf{E} & \cellcolor{cW}\textbf{E} & \cellcolor{cW}\textbf{E} &  &  &  & \cellcolor{cD}\textbf{C} & \cellcolor{cD}\textbf{C} &  &  &  &  & \cellcolor{cDiag} &  &  \\
\arrayrulecolor{catP}\hline
\textcolor{catP}{P4} & \nbox{Lack of deterrence and enforcement} & \cellcolor{rev3}{\scriptsize Reversible} &  &  &  &  &  &  & \cellcolor{cD}\textbf{C} &  &  &  &  &  & \cellcolor{cW}\textbf{E} & \cellcolor{cDiag} &  \\
\arrayrulecolor{catP}\hline
\textcolor{catP}{P5} & \nbox{Geopolitical changes} & \cellcolor{rev2}{\scriptsize Difficult} & \cellcolor{cW}\textbf{E} & \cellcolor{cW}\textbf{E} &  &  &  &  &  & \cellcolor{cD}\textbf{C} &  &  &  &  & \cellcolor{cW}\textbf{E} & \cellcolor{cW}\textbf{E} & \cellcolor{cDiag} \\
\arrayrulecolor{black}\hline
\arrayrulecolor{black}
\end{tabular}
\vspace{0.2cm}
\caption{Dependencies among the pathways to losing the option to restrain AI development.\\ Cells are read row-to-column: \colorbox{cD}{\textbf{C}} = the row  may directly \emph{cause} the column; \colorbox{cW}{\textbf{E}} = the row \emph{exacerbates} the effects of the column. Reading across a row lets one see the effects of a pathway. For example, if existing hardware cannot be tracked (H1), then this will exacerbate the effects of distributed training and algorithmic progress because there will be more unmonitored hardware to train AIs with. Reading down a column lets one see possible causes or exacerbating factors for a pathway. For example, we can see that many pathways make verification more politically costly and make states more likely to refuse international verification (P3).\\
Pathways are also classified by their reversibility: Almost impossible to reverse, difficult, or reversible. }
\label{tab:pathway-dependencies}
\end{table*}

\subsection{Hardware governance failures}

The regime outlined above depends on the ability to locate, consolidate and monitor AI hardware~\citep{scher2025international}. It becomes infeasible if states cannot track existing hardware, cannot track new hardware, or cannot trust monitoring and verification methods to withstand attempts at subversion.

\subsubsection{Inability to track existing hardware}

AI hardware may proliferate too widely for governments to locate and consolidate a sufficiently large share. Smuggling is a particular concern~\citep{grunewald2023aichipsmuggling, fist2023preventing, scannell2026supermicro}, as a lack of formal records may render smuggled chips invisible to financial intelligence and supply chain records that would otherwise enable tracking. The problem is compounded if states that refuse to comply with international tracking acquire substantial AI hardware. A small number of non-complying states might be manageable through diplomatic or economic pressure, or the threat of force. But sufficient untracked hardware in such states could render tracking infeasible. 

Governments themselves may contribute to this issue. States anticipating a future global prohibition on advanced AI development might preemptively stockpile compute, in order to retain unmonitored AI capabilities after restrictions take effect. States might also have large quantities of AI chips in sensitive military data centers, and may not be willing for these chips to be part of tracking efforts. 

An additional challenge is mutual verification: states must be able to verify to their rivals that their borders do not contain significant untracked compute. Even if a state honestly tracks and consolidates its own hardware, other states must be able to verify this. In particular, this could be an issue with sensitive military sites that may be housing AI chips. 

Currently, the vast majority of AI chips weighted by performance are concentrated into relatively few large data centers, which would make tracking much easier~\citep{pilz2025trends, EpochAIDatacenters2025}. This simplifies tracking, but it would still require international coordination. In the absence of deliberate intervention, AI hardware will become more proliferated, at least in terms of absolute numbers of chips. Even if a large share remains easy to track, a sufficient absolute quantity of untracked chips may  undermine a restraint regime (e.g., sufficient to perform a large, unmonitored training run). 

Widespread proliferation of AI hardware might be effectively irreversible. The placement of many AI chips into hidden facilities would be more reversible, but would require large political buy-in to remediate. 

\subsubsection{Inability to track new hardware production}

The AI chip supply chain is currently extremely concentrated~\citep{thadani2023mapping}. A small number of fabrication facilities produce virtually all high-end AI chips. Inputs and equipment---such as high-bandwidth memory and EUV lithography machines---are also concentrated among a handful of suppliers. This concentration makes monitoring and controlling the production of new AI chips tractable, because few facilities are involved. If the supply chain became more diffuse, tracking new chip production would become substantially harder.   

A speculative potential development is the emergence of a new chip manufacturing paradigm that allows AI chips to be produced without the massive fabrication facilities currently required. This does not appear likely in the near term, as current chip manufacturing is among the most complex industrial processes ever developed and, historically, new semiconductor manufacturing processes take decades to mature~\citep{miller2022chip}.  

Overall, tracking new AI chip production and maintaining supply chain concentration appears considerably easier than tracking existing chips that have already been sold and distributed.

\subsubsection{Hardware monitoring subversion}

Once AI chips have been consolidated into monitorable data centers, the AI chips must then actually be monitored to ensure they are not being misused.   

Ideally, monitoring will be able to distinguish training from inference without being overly invasive~\citep{baker2025verifying, scher_mechanisms_2024, heim_trusted_2024}. Some training runs should be permitted (e.g., with small compute budgets or aimed at narrow, safe domains). In this case, monitoring must be able to verify that only approved workloads are running.   

However, methods could be developed to subvert chip monitoring, or the perceived probability of subversion may be high enough to effectively undermine mutual verification efforts.

In cases where chip monitoring measures are not trusted, states may rely on more costly or invasive measures. States could require and provide full visibility into which workloads are run on AI chips. Or as a last resort, states might simply power off or physically disconnect their AI chips and allow inspectors from rival states to verify that the hardware is depowered. This would eliminate the economic benefits of continued inference, making restraint significantly more costly.

\subsection{Erosion of training threshold effectiveness}

Training thresholds will only be effective if two conditions hold: advanced AI models require large, monitored clusters to train, and models trainable below the compute threshold remain non-dangerous. Several developments could undermine one or both of these conditions.

\subsubsection{Distributed training}

Distributed training is one of the main ways a catastrophically dangerous AI model could be trained using small, unmonitored clusters, while still requiring large overall compute~\citep{krys2025distributed}. 

If distributed training techniques advance to the point where catastrophically dangerous models could be trained across individually permitted clusters, the restraint regime would need additional monitoring or restrictions.  

There are several possible responses. It might be sufficient to enact a maximum memory requirement for unmonitored clusters~\citep{rahman2026catching}. If this was insufficient, the maximum size of unmonitored clusters could be lowered, but this would increase the difficulty and intrusiveness of compute tracking. Internet service providers (ISPs) could be enlisted to look for suspicious data transfer patterns that might indicate distributed training operations, although such attempts could possibly be subverted by competent adversaries. Alternatively, intelligence agencies could be tasked with uncovering such operations directly.

\subsubsection{Continued algorithmic progress}

Over time, algorithmic efficiency improvements reduce the compute required to achieve a given level of AI capability. Historically, the rate of algorithmic progress has been quite fast, with estimates ranging from 3$\times$ to 60$\times$ per year for the reduction in compute needed to reach a level of AI capabilities~\citep{ho_algorithmic_2024, epoch2026theleastunderstooddriverofaiprogress, scher2025catchup}.

Increased training efficiency leads to two problems. First, for training on monitored clusters, if catastrophically dangerous AI models can be trained using an amount of compute below the training compute threshold, then the threshold must be lowered.  Second, as the compute required to train catastrophically dangerous models decreases, it will be possible to train these models on small, unmonitored clusters. If algorithmic progress continues unabated, at some point compute-based governance would likely become infeasible, for example if a single consumer GPU were sufficient to train a catastrophically dangerous model. 

\subsubsection{New paradigms and inference scaling}

More speculatively, the development of new, dramatically more compute-efficient AI paradigms could undermine training compute thresholds, potentially rendering the entire approach obsolete.   

There is a spectrum between continued business-as-usual algorithmic progress and a genuinely new paradigm. A relevant case along this spectrum is inference scaling, where models can be run for longer to increase their capabilities, as with current `reasoning' models. Inference scaling is another axis of scaling beyond training compute. If inference scaling can increase model capabilities to dangerous levels, this may necessitate inference monitoring (see Section~\ref{sec:Development_dangerous_models}). However, if significant compute is required for dangerous amounts of inference scaling, it may only be feasible on large clusters that are already subject to monitoring.

\subsubsection{Research restrictions fail}

The developments described above–advances in distributed training and algorithmic progress–may not have occurred by the time a restraint regime is established. But after such a regime is established, new methods could still be developed that erode compute thresholds. This may necessitate restrictions on AI research. Ideally these restrictions would be narrowly targeted at research that increases general AI capabilities, in order to maximize the benefits from AI that is not catastrophically dangerous. The exact scope of restrictions on AI research is not the focus of this work; however \emph{some} restrictions on AI research would likely be needed if governments wanted to restrain AI progress for more than a few months~\citep{ho_algorithmic_2024, scher2025catchup, epoch2026theleastunderstooddriverofaiprogress}.    

Enforcing and internationally verifying research restrictions may be challenging. One approach is to track researchers and conduct regular interviews about their work. This would be undermined if researchers relocated to states that are difficult to monitor or that refuse to cooperate with tracking requirements.   

Additionally, verifying the absence of state-run AI research programs could pose a similar challenge to hardware inspections. States that are unwilling to grant inspectors access to military data centers may also be unwilling to grant access to classified AI research programs. This is an issue whether or not state-run AI research programs exist, as other states will need to be able to verify they do not.  

A potential advantage is that the population of researchers capable of meaningfully advancing the frontier is currently relatively small~\citep[p.~38]{scher2025international}. This could make it challenging for a government to recruit top researchers into a secret program without this being noticed.

If AI systems themselves are able to automate AI research, then restrictions focused on human researchers would be far less effective. Therefore, the development of such AI systems would make restraint far more difficult, as discussed in the following section.

\subsection{Catastrophically dangerous models}
\label{sec:Development_dangerous_models}
The above pathways (hardware governance failures and erosion of training threshold effectiveness) focus on losing the ability to restrain the \textit{training} of catastrophically dangerous AI models, because restraining training is much easier than restraining inference. If catastrophically dangerous AI models are nevertheless developed, restraint would require monitoring or controlling inference.

\subsubsection{Development of catastrophically dangerous models}
\label{sec:dangerousmodel_dev}
If a catastrophically dangerous AI model is trained (but not necessarily released), then inference monitoring will be necessary to ensure the model is not being used. As discussed above, “catastrophically dangerous” here refers to models that are directly capable of causing large-scale harm, models that can sufficiently automate AI research to lead to a directly catastrophically dangerous model, or models that could be modified into either. 

Such a model would be highly capable, creating strong incentives to develop it despite the risks. The incentives might be especially strong to develop models that can automate AI research, as the danger from automating AI research may be less apparent than the dangers posed by more immediately capable systems. There would likely also be disagreement about how much risk using the model would entail. 

Simply deleting the model would not resolve the problem. Once a model has been trained and its existence is known, it could be infeasible to verify to other states that all copies have been deleted, even if they had been. AI models are software and can easily have their code and weights copied~\citep{brown2026aggressive}; the weights of the top open models total around 1 TB~\citep{kimi2025k25,xu2026deepseek}.  This is compounded by standard training practices: frontier models are typically trained using many instances of the model weights on different hardware, with numerous checkpoints saved throughout. That is, the ordinary training process already involves creating many copies of a model, across many distinct chips. Even if tamper-proof logs were maintained throughout training, model weights could potentially be extracted via hardware side-channel attacks~\citep{joud2022practical}. 
This isn’t merely an external threat. An AI developer may themselves attempt to extract model weights from secure hardware, and would find it much easier to do so than an external party. The verifiable deletion of AI model weights is an unsolved problem.

Therefore the creation of a catastrophically dangerous AI model should likely trigger, at a minimum, inference monitoring of the AI developer, as well as continued surveillance to ensure that model weights were not being exfiltrated. Even this may not be sufficient, as the model weights may have been exfiltrated prior to surveillance beginning. 

\subsubsection{Release of catastrophically dangerous models}

If the weights of a catastrophically dangerous model are released to the public or covertly exfiltrated, restraint becomes even harder. If a released model is small enough to run on an unmonitored cluster, it could become effectively impossible to prevent people from running inference. Given the relatively small size of frontier AI models, it is very likely that a catastrophically dangerous model could run on an unmonitored cluster.

If model weights were shared with a limited group rather than posted publicly and downloaded by thousands, authorities might be able to intervene to stop further spread. But verifying that the weights had not been copied further would be extremely difficult.

\subsubsection{Release of autonomous AI models}

Future AI systems may be able to run autonomously, without reliance on humans. Such systems might be capable of acquiring compute resources, copying their weights to new servers, and adapting to new challenges like attempts to shut them down~\citep{the-rogue-replication-threat-model}. This set of capabilities is referred to as autonomous replication and adaptation (ARA)~\citep{kinniment2023evaluating}. Not all ARA-capable systems would necessarily be dangerous; some might spread widely while remaining far from capable of causing large-scale harm. But if a catastrophically dangerous ARA-capable system were released, shutting it down could prove extraordinarily difficult. One report from RAND explores extreme countermeasures like high-altitude electromagnetic pulses to disrupt ground-based electronic infrastructure, or attempting to shut down global internet infrastructure~\citep{PE-A4361-1}. Even these extreme measures might prove insufficient, for example due to data centers with shielding against electromagnetic pulses~\citep{davis2022emp}.

\subsection{Risks to political feasibility}

The above pathways describe developments that would make restraint more technically difficult or costly, even given political consensus that restraint is needed. Here we address a different class of threat: states of the world that would make restraint politically difficult to pursue in the first place. These cover domestic and international dynamics that would make restraint politically difficult.

\subsubsection{Societal dependence and public support}

As AI systems become more capable, society may develop a widespread dependence on them~\citep{kulveit2025gradual, sharma2026s, drago2025intelligence}. As AI becomes more integrated into the economy, military, and daily life, the most salient costs of restraining further development will grow–even though these immediate costs would likely be outweighed by the risks from continuing development. Large investments will be based on the assumption of continued capabilities progress, such as with current data center buildouts~\citep{epoch2025whatyouneedtoknowaboutaidatacenters}. 

Beyond this dependence, there may be widespread public support for continued AI development. AI systems might help automate dangerous work, contribute to scientific or medical breakthroughs, or become part of people's personal lives as AI companions and assistants~\citep{bernardi2025friends}.  Alternatively, AI companies might claim such benefits will come with future AI systems, even if they've yet to materialize. In either case, governments may face popular opposition to restraint. However, this support might be counteracted by negative impacts such as job displacement or the misuse of AI systems by malicious actors.

Ideally, a restraint regime would allow continued inference on existing AI systems, reducing the material and political costs of restraint.

\subsubsection{AI company influence}

AI companies themselves will likely grow in political and economic power. Company leadership is selected for believing in the upside of building advanced AI, and will likely perceive enormous personal reward from continued AI development---even if this development is risky for the world more broadly. These companies can be expected to lobby against restraint, similarly to how they lobby against current attempts at regulation~\citep{leadingthefuture2025launch, publiccitizen2026lobbyists, liu2026builders}. This problem could be intensified by AI-enabled persuasion and propaganda~\citep{hackenburg2025levers, kowal2025s, rogiers2024persuasion, costello2026large}: AI companies may use their own AI systems to influence public opinion in favor of continued development. There has already been one documented case of an AI company apparently funding an AI-generated news site focused on discrediting critics~\citep{johnston2026acutus}.

\subsubsection{States refuse international verification}

As discussed throughout, restraint on AI development must be global and internationally verifiable. Catastrophically dangerous AI developed in any jurisdiction would pose a global threat, and states will likely only agree to prolonged restraint if they are confident their geopolitical rivals are doing the same.

Therefore, a major obstacle to restraint is that intrusive verification may be politically unacceptable to major powers. A lack of trust between major powers could make verification politically unacceptable, particularly for sensitive sites that may house data centers or state-run AI research programs. However, the START treaties offer some precedent for this style of verification: the U.S. and USSR agreed to invasive monitoring of nuclear sites despite deep mutual distrust~\citep{talbottDeadlyGambitsReagan1985, bennettRussianNegotiatingStrategy1997}. World leaders are unlikely to accept such verification measures today, but may be more likely to if they believe they are necessary to avoid a race to catastrophe.  

Less intrusive methods might partially address verification of sensitive sites without requiring full physical inspections. Satellite imagery, power infrastructure analysis, and supply chain paper trails could help verify that chips have not been diverted to unmonitored facilities.

\subsubsection{Lack of deterrence and enforcement}

An international restraint regime likely requires credible deterrence. States must believe that violations will be detected and met with consequences. In extreme cases, it might be necessary to disrupt AI development via military force~\citep{hendrycks_superintelligence_2025}, akin to the most extreme nuclear counter-proliferation efforts. Leaders might be unwilling to commit to such actions, leading to insufficient deterrence.  

However, if leaders were sufficiently concerned to restrain their own AI development, they would likely also be concerned enough to follow through with enforcement.

\subsubsection{Geopolitical changes}

Currently, the U.S. and China are both the primary actors in AI development and the dominant geopolitical powers. This means a restraint regime could plausibly be built around a U.S.--China agreement, with each bringing in their respective allies.

Two developments could make this harder. First, if AI development becomes more geographically distributed, coordination will be more difficult, because there would be more individual actors to appease. Second, the world may become more multipolar, with states less clearly aligned with either the U.S. or China. This would break down the assumption that the two major powers could simply bring their allies into the fold.

A direct military conflict between the U.S. and China would also make even basic diplomatic engagement infeasible, let alone negotiation over a mutual verification regime. A prolonged conflict could entirely prevent international coordination on AI. Even if leaders on both sides came to believe that AI posed an existential threat, rebuilding the diplomatic channels needed for mutual restraint from a standpoint of active hostility would be extraordinarily difficult. Worse, active conflict could increase the pressure to develop more advanced AI, as either side may  believe it would offer a decisive military advantage.

\section{Policy implications}

Without deliberate action, governments may lose the option to restrain AI development, even if there is widespread agreement at the time that restraint is necessary. In some scenarios, restraint may become effectively impossible; in others, it may remain possible but at great economic, political, or social cost. There are policy measures that could be implemented today that would help to preserve the option of later restraint. The window to implement these measures is closing, because many of the developments described are irreversible once they have happened. 

\paragraph{Is the window actually closing?} There are several trends that, if they were to continue, may effectively remove governments' ability to restrain AI development, rather than just make this more costly. For example:
\begin{itemize}
    \item \textbf{Hardware proliferation}, via smuggling~\citep{grunewald2025countering} and legitimate channels~\citep{reuters2026h200}, would make it impossible to ensure there are not large quantities of untracked chips. Even though the majority of AI chips are in U.S. data centers, the absolute number of AI chips in smaller countries is growing rapidly~\citep{pilz2025trends}. 
    \item \textbf{Advances in distributed training} would allow large training runs to continue on small untracked compute~\citep{krys2025distributed, epoch2025howfarcandecentralizedtrainingovertheinternetscale}. Frontier AI companies~\citep{charles2026communication,douillard2026decoupled} and smaller groups~\citep{lidin2026covenant} continue to publish research into efficient distributed training. 
    \item \textbf{AI companies are racing to automate AI R\&D}, which may directly lead to catastrophically dangerous models and remove the ability to restrict research~\citep{field2026ai,favaro2026recursive}.
    \item \textbf{AI companies are building increasingly capable models}, which could soon be catastrophically dangerous~\citep{kokotajlo2025ai2027}, and which, if developed, could not be verifiably deleted.
\end{itemize}

\paragraph{Points of no return} One unfortunate factor in deciding when to act is that there are not clear “points of no return” for the pathways described above. Nobody currently knows exactly how much hardware proliferation would be too much, what capabilities an AI system will have before it is trained, or how to measure whether a particular AI system is capable enough to autonomously catalyze catastrophically dangerous AI development. We may only know once we have passed a threshold that we couldn’t see in advance. Defining and measuring such thresholds would require a much better understanding of AI development than the field currently has~\citep{barnett_what_2024}. Therefore we suggest a conservative approach for policymakers, acting early to preserve governance options, rather than waiting for clearer signals that may arrive too late.

\subsection{Track existing AI hardware}

One of the clearest ways governments could lose the option to restrain AI development is if advanced AI hardware becomes too difficult to track. States should therefore seek to identify stockpiles of AI chips (within their jurisdictions and abroad), while strengthening export-control enforcement and counter-smuggling efforts to limit proliferation of untracked AI hardware. 

For the next generation of AI chips, on-chip location-tracking mechanisms could be used to verify where chips are physically located~\cite{brass2024location, aarne2024secure, harack2025verification, scher_mechanisms_2024, kulp2024hardwareenabled}. Governments could require that new AI chips are equipped with these mechanisms, and that the mechanisms are always-on and tamper-resistant---such that chips must securely attest to their location in order to function. The always-on requirement is important for restraint: a mechanism that can be disabled cannot verify that a chip is not being used in an unauthorized location.

The continued distribution of AI chips to many small actors–as opposed to a smaller number of easily trackable hyperscalers–would make future consolidation significantly more challenging. States could mitigate this by limiting direct sales of AI chips to small buyers and instead promoting remote cloud access as the primary way to use AI hardware~\citep{heim_governing_2024}. This would keep AI hardware physically consolidated in a small number of large, trackable data centers while still providing broad access.

\subsection{Keep the chip supply chain concentrated}

The current concentration of the semiconductor supply chain is a governance asset that should be actively preserved. Export controls on semiconductor manufacturing equipment can help prevent the diffusion of fabrication capabilities to additional states. A narrow supply chain makes it easier to track new chips as they enter the market, since they originate from a limited number of sources. It also means that fewer facilities would need to be monitored under a restraint regime. 

At a minimum, governments should only allow the proliferation of semiconductor manufacturing equipment into jurisdictions likely to comply with some future verification regime.

\subsection{Develop chip monitoring mechanisms}

In order to preserve access to safe inference under a restraint regime, mechanisms are needed to distinguish training from inference, and potentially to verify that only permitted training workloads are running~\citep{petrie2024near, baker2025verifying, aarne2024secure,  kulp2024hardwareenabled}. Governments should incentivize research in this area, by directing government research labs, funding academic and private groups, or mandating these measures as a requirement for chip manufacturers. 

There is early work in this direction, but these mechanisms need to advance beyond prototypes and demonstrations; they must be robust enough to withstand state-level attacks. Hardware-based monitoring can be complemented by physical measures such as on-site inspectors or tamper-proof cameras to ensure chips are not physically modified to circumvent monitoring.

\subsection{Track AI capabilities}

Governments need to know when they should restrain AI development. This should be before directly catastrophically dangerous AI systems are developed, and before AI systems are developed that would require inference monitoring, such as systems capable of automating AI research. 

To achieve this, governments should establish information-sharing and other transparency mechanisms with domestic AI developers~\citep{belfield_what_2024}, with particular attention to progress in AI research automation and other dangerous capabilities. At a minimum, these transparency mechanisms should include mandatory reporting of AI capabilities, incidents, and hardware stockpiles, as well as whistleblower protections for employees. Further options include: third-party audits, interview programs~\citep{wasil_understanding_2024}, and continuous resident inspectors modeled after the Nuclear Regulatory Commission.

This transparency also helps governments to become aware of dangerous models before they are released or further developed. Without visibility into AI development, a catastrophically dangerous model could be trained and released without government officials even knowing it existed. 

Ideally, governments would have accurate threat models and corresponding benchmarks to track relevant AI capabilities. Unfortunately, work on AI threat modeling is nascent~\citep{campos2025frontier} and benchmarks often fail to capture important AI capabilities~\citep{barnett_what_2024, barnett_declare_2024, mukobi2024reasons}. This means it is challenging to translate from benchmark results to a level of risk, and thus even harder to set clear red lines around AI capabilities. 

Governments should also track AI progress in other countries, both to assess global risk levels and to detect when rivals may be approaching dangerous capability thresholds.

\subsection{Engage in diplomatic efforts}

International restraint will require a minimum level of trust between major powers sufficient to agree to a mutual verification regime. U.S.--China diplomacy is a key priority, as these are the two dominant actors in both AI development and geopolitics.

One avenue for diplomacy is reciprocal sharing of information about AI development; for example, sharing information on the locations of major AI data centers and updates on internal AI capabilities. Such information sharing helps with two goals: it builds the diplomatic foundation for a future verification regime, and it helps states assess whether their rivals are approaching the development of catastrophically dangerous AI systems. This information would likely be important for deciding when to mutually restrain AI development. 

\section{Conclusion}
Whether governments will retain the ability to restrain advanced AI development depends on choices made today. In this paper, we have described a possible regime for restraint, built around training compute thresholds, hardware consolidation, research restrictions, and international verification. Following this, we identify four categories of developments that could render such a regime infeasible: hardware governance failures, the erosion of training threshold effectiveness, the development of catastrophically dangerous models, and various risks to political feasibility. Many of the developments within these categories are difficult or impossible to reverse once they occur.

We make various policy recommendations: tracking AI hardware, preserving supply chain concentration, investing in chip monitoring, building visibility into AI capabilities, and engaging in diplomacy. These recommendations do not constitute a halt on AI development; they are initial infrastructure that would make restraint possible if it were ever judged necessary. 

A core difficulty is that there are no reliable indicators before a point of no return. This suggests a conservative approach: governments should act now to preserve future optionality rather than wait for evidence that may arrive too late.

\bibliography{refs_pb}
\bibliographystyle{icml2026}
\end{document}